\documentclass[12pt,preprint]{aastex}

\def\magarc{\ mag arcsec$^{-2}$}
\def\gtsim{~\rlap{\lower -0.5ex\hbox{$>$}}{\lower 0.5ex\hbox{$\sim\,$}}}
\def\ltsim{~\rlap{\lower -0.5ex\hbox{$<$}}{\lower 0.5ex\hbox{$\sim\,$}}}
\def\arcmin{\hbox{$^{\prime}$}\,}

\def	\beq	{\begin{equation}}
\def	\cm	{\,{\rm cm}}

\def	\eeq	{\end{equation}}

\def	\etal	{{\it et al.}}

\def	\gtsim	{\gtrsim}		 

\def	\incl	{{i}}

\def	\kms	{\,{\rm km~s}^{-1}}

\def	\ltsim	{\lesssim}		 

\def	\sr	{\,{\rm sr}}

\def\gta{\;\lower 0.5ex\hbox{$\buildrel > \over \sim\ $}}
\def\lta{\;\lower 0.5ex\hbox{$\buildrel < \over \sim\ $}}
\def\arcsec{$^{\prime\prime}$}
\def\arcmin{$^{\prime}$}
\def\deg{$^{\circ}$}
\def\etal{{\it\ et al.}}
\def\ie{{\it i.e.}}
\def\eg{{\it e.g.}}

\def\kms{\ km s$^{-1}$}

\begin{document}

\title{NGC 300: an extremely faint, outer stellar disk observed to 10 scale lengths}

\smallskip
\author{J. Bland-Hawthorn}
\affil{Anglo-Australian Observatory, PO Box 296, Epping, NSW 2121, Australia}
\author{M. Vlaji\'{c}}
\affil{Anglo-Australian Observatory, PO Box 296, Epping, NSW 2121, Australia}
\author{K.C. Freeman}
\affil{Mount Stromlo Observatory, Private Bag, Woden, ACT 2611, Australia}
\author{B.T. Draine}
\affil{Princeton University Observatory, Princeton University, Peyton Hall, Princeton, NJ 08544-1001, USA}

\begin{abstract}
We have used the Gemini Multi-object Spectrograph (GMOS) on the
Gemini South 8m telescope in exceptional conditions (0.6\arcsec\
FWHM seeing) to observe the outer stellar disk of the Sculptor group
galaxy NGC 300 at two locations.  At our point source detection
threshold of $r' = 27.0$ (3$\sigma$) mag, we trace the stellar disk
out to a radius of 24\arcmin, or $2.2\:R_{25}$ where $R_{25}$ is
the 25\magarc\ isophotal radius.  This corresponds to about 10 scale
lengths in this low-luminosity spiral (${\rm M}_B = -18.6$), or
about 14.4 kpc at a cepheid distance of 2.0 $\pm$ 0.07 Mpc.  The
background galaxy counts are derived in the outermost field, and
these are within 10\% of the mean survey counts from both Hubble
Deep Fields.  The luminosity profile is well described by a nucleus
plus a simple exponential profile out to 10 optical scale lengths.
We reach an effective surface brightness of $\mu_{r^{'}} = 30.5$\magarc\
($2\sigma$) at 55\% completeness which doubles the known radial
extent of the optical disk.  These levels are exceedingly faint in
the sense that the equivalent surface brightness in $B$ or $V$ is
about 32\magarc. We find no evidence for truncation of the stellar
disk. Only star counts can be used to reliably trace the disk to
such faint levels, since surface photometry is ultimately limited
by nonstellar sources of radiation.  In the Appendix, we derive the
expected surface brightness of one such source: dust scattering of
starlight in the outer disk.

\end{abstract}

\keywords{galaxies: individual (NGC~300) $-$ galaxies: photometry $-$
galaxies: formation, evolution, dynamics, structure $-$ ISM: dust}

\bigskip
\begin{figure*}
\caption{
Wide field image (30\arcmin$\times$20\arcmin) of NGC 300 from the 
Digitized Sky Survey (North is at the top and East is left).  
The insets show the summed data from the two 
GMOS outer fields. The GMOS field of view is $5'\!\!.5$. The fields 
are located at a radius of 15.47\arcmin\ (Field 1) and 21.14\arcmin\
(Field 2).
}
\label{DSS}
\end{figure*}
\bigskip

\section{Introduction}
The structure of the outer parts of galactic disks is central to
our understanding the formation and evolution of spiral galaxies.
The surface brightness profile of a spiral galaxy is a first order
signature of the formation process (Freeman \& Bland-Hawthorn 2002;
van der Kruit 1987; Ferguson \& Clarke 2001).  By measuring scale lengths
for galaxy samples of different redshifts, it is for example possible
to directly address the evolution of galactic disks (Labb\'{e}\etal\
2003).

To a first approximation, the radial light distribution of disks
is well described by an exponential decline (Freeman 1970).  If the
exponential is due to dynamical evolution (Lin \& Pringle 1987), it is
important to establish out to what radius this process took place.
It is possible that departures from an exponential will show up at
faint levels, reflecting new forms of dynamical evolution or different
angular momentum of the outermost accreted material.

Previous studies have shown that the exponential light distribution
breaks smoothly at a well-defined `break radius' in some galaxies
(Pohlen et al.~2002; de Grijs et al.~2001). In the optical, this
occurs at galactocentric distances of typically 3-5 exponential
scale lengths. In principle, this break radius could be used as an
intrinsic parameter measuring the size of a galactic disk.  However,
despite the fact that these breaks have been known for more than
20 years (van der Kruit 1979), we do not have a secure understanding
for the origin of the break, so their application as a fundamental
disk characteristic is therefore not yet possible.

Beyond the break radius, stars are detected albeit with a steeply
declining luminosity profile (\eg\ Kregel, van der Kruit \& de Grijs
2002). To date, the outer edge of the stellar disk (`truncation
radius') has not been detected in any galaxy, largely because diffuse
light photometry is unreliable at such faint magnitude limits. We
note that in two previous papers, optical disks are seen to extend
to 10 optical scale lengths. In their study of NGC~5383, Barton \&
Thompson (1997) reach down to almost $\mu_V \approx 30$\magarc\
(1$\sigma$).  In a multiband study of NGC~4123, Weiner\etal\ (2001)
reach 29, 29, 28\magarc\ (1$\sigma$) in $B$, $V$ and $I$ respectively.
However, we point out that both of these galaxies are strongly
barred and have peculiar morphologies.

We choose to focus our first study of outer disks on the Sculptor
group for several reasons.  It is the closest galaxy group to the
Local Group with a high proportion of disk systems. Furthermore,
the group has not undergone much dynamical evolution $-$ the galaxies
are more or less isolated. Their dark haloes are thought to extend
far beyond the bright disks. In principle, accretion onto the halo
could give rise to disks which are much larger than normally assumed.

The layout of the paper is as follows. Our motivation for targetting
NGC~300 is discussed in \S 2. We briefly discuss surface photometry
vs. star counts in \S 3 and show that the latter is the only way
to get below 30\magarc\ reliably, an issue we return to in the Appendix.  The
observations and reductions under IRAF are discussed in \S 4, while
details of the photometric calibration are given in \S 5. The results
are presented in \S 6 and these are discussed in \S 7.

\section{Object selection}

We chose NGC~300 as a relatively undisturbed member of the closest
galaxy group, Sculptor.  NGC~300 lies at the nearside of a distribution
which is elongated along the line of sight (Jerjen, Freeman \&
Binggeli 1998).  There have been various distance estimates based
on cepheid variables (Freedman\etal\ 2001; Gieren\etal\ 2004) and
the tip of the red giant branch (Butler\etal\ 2004). We adopt a
weighted mean of 26.5$\pm$0.1 for the distance modulus, equivalent
to 2.0$\pm$0.1 Mpc.  We show below that our photometry reaches down
to an intrinsic stellar magnitude of $M_r^{'} = 0.5$ (3$\sigma$) which
includes some of the giant branch and much of the horizontal branch
in a moderately metal poor ([Fe/H]$\approx$-1) population.

The late type spiral galaxy NGC~300 is ideal for our proposed study.
It is an almost pure disk galaxy (bulge light $<$ 2\%) and has a
mild inclination ($i=42$\deg).  The galaxy lies at high galactic
latitude ($b\approx -79$\deg), and therefore has low foreground
reddening of E(B-V)=0.013 mag.  NGC 300 shares important similarities
with M33 in the northern hemisphere (Blair \& Long 1997).  The
optical disk scale length has been measured as 2.17 kpc in $B_J$
(Carignan 1985; scaled to a distance of 2.0 Mpc) and 1.47 kpc in
$I$ (Kim\etal\ 2004). We note that these scale lengths would indicate
that the colour of the outer disk is likely to be very blue, but
this has not been measured to date (see \S 7). Carignan's (1985;
Fig.~8) photometric study reveals that the outermost contours are
well behaved, with minimal asymmetry.

\section{Surface photometry vs. star counts}

There are two basic methods for detecting the faint outer reaches
of galaxies: deep diffuse light imaging (\eg\ de~Vaucouleurs \&
Capaccioli 1979), and faint star counts (\eg\ Pritchet \& van den
Bergh 1994).  

Surface photometry of face-on galaxies provides radial information
which is more easily interpreted, but is limited by the brightness
of the night sky.  Edge-on galaxies are easiest to search for optical
breaks and truncations because of the enhanced surface brightness
from the line-of-sight integration.  However, with edge-on systems
one cannot be sure that the surface photometry is not affected by
smaller-scale structures in the disk, like faint spiral arms for
example, which might mimic the effects of truncation (Pohlen\etal\
2004). Dust extinction is also more severe in edge-on systems although
less so at large radius.

While diffuse light imaging is a direct method, there are many
technical difficulties at levels below $\mu_R  \approx 27$\magarc\
(\eg\ Pohlen\etal\ 2002). These include scattered light internal to
the optics, difficulties of flatfielding data (Pohlen\etal\ 2002),
zodiacal light (Bernstein\etal\ 2002), and diffuse Galactic light
(Haikala \& Mattila 1995), a combination of dust-scattered starlight,
dust luminescence,
and emission from cool gas.  Over the sky, there are also variations
in Galactic cirrus (Schlegel, Finkbeiner \& Davis 2000), and warm
ionized gas due to the Reynolds layer (Reynolds 1992).  At fainter
levels, in the range $\mu_R \approx 30-32$\magarc, the outer HI
envelope may have (i) a sufficient column of dust to scatter
detectable levels of starlight from the optical disk (see Appendix),
or (ii) sufficient levels of ionization due to a local or global
radiation field to produce fluorescent emission (Maloney 1993;
Maloney \& Bland-Hawthorn 2001).

\bigskip
\begin{figure*}
\caption{
Rejection masks for the GMOS fields. The elliptic arcs illustrate
how the star counts are binned in azimuth.  The major axis of the
annuli passes through the middle of Field 1. For consistency, all
radial profiles are presented in terms of the major axis distance
from the galactic nucleus (\ie\ defined by the major axis radius
of the elliptic annulus). The region eclipsed by the on-instrument
wavefront sensor can be seen in the top right hand corner of both
fields. The $(x,y)$ axes are in units of CCD pixels.
} \label{MASK} \end{figure*} \bigskip

A more powerful method is to count individual stars in the old
stellar population.  The technique is particularly effective since
good seeing allows contamination from background galaxies and
foreground stars to be properly evaluated.  In their study of
the M31 halo ($(m-M)_o \approx 24.5$), Pritchet \& van den Bergh
(1994) achieved 50\% completeness for point sources down to $V=24.5$
in 1\arcsec\ seeing. This corresponds to an `effective' surface
brightness limit of $\mu_V = 29$\magarc\ after correcting for
foreground contamination and the stellar disk contribution from
unresolved stars.  With their 4m ground-based observations, Pritchet
\& van den Bergh reach to $M_V \approx 0$, corresponding to roughly
40\% of starlight from the old stellar population.  More recently,
very deep Hubble Space Telescope observations reach $m_I = 31$ at
50\% completeness in a relatively small 3\arcmin\ field (Brown\etal\
2003).

With stellar photometry, we can trace the disk to much fainter
effective surface brightnesses by resolving individual stars.  This
is the motivation for our deep GMOS observations.  It is however
crucial to establish the magnitude completeness limit and the
fraction of missing light below the detection limit, which we
discuss.

\bigskip
\begin{figure*}
\caption{
A $25''\times 25''$ subset from the centre of Field 1 to demonstrate
the quality of the {\sl daophot} photometric analysis. The images show
(a) before source subtraction;
(b) with $5 \sigma$ sources subtracted;
(c) with $3 \sigma$ sources subtracted.
The psf is mildly elliptic ($10-15\%$) and
$0^{''}\!\!\!.60$ along the long axis.
}
\label{PSF}
\end{figure*}
\bigskip

\section{Observations and Data Reduction}

Deep $r^{\prime}$ images of NGC 300 were obtained in two outer
locations using Gemini Multi-Object Spectrograph (GMOS) on Gemini
South during semester 2003B.  The two GMOS fields lie close to the
major axis of NGC 300 and span a major axis distance range of 7 to
16 kpc.  Our science goal was to track the radial profile of the
galaxy as far as possible which required the best possible observing
conditions.  Of the available filter set, the $r'$ filter 
(Fukugita\etal\ 1996) is the
most efficient since (i) this is where the performance of GMOS peaks
(Hook\etal\ 2004), (ii) the resolved stellar population will probably
be dominated by red giants, and (iii) the sky in $r^{\prime}$ is
significantly darker than in redder bands like $i^{\prime}$.

\begin{center}
\begin{table}
\begin{tabular}{l l l l l}
 & $\alpha$ (J2000) & $\delta$ (J2000) & Date (2003) & Exposure (s) \\ \\
Field 1 & $00^{h}\,55^{m}\,50^{s}$ & $-37^{\circ}\,47'\,20''$ & Dec 20-22 & $9 \times 900$\\
Field 2 & $00^{h}\,56^{m}\,15^{s}$ & $-37^{\circ}\,46'\,49''$ & Dec 23-26 & $9 \times 900$\\
Landolt 95 & $03^{h}\,53^{m}\,01^{s}$ & $+00^{\circ}\,00'\,13''$ & Dec 20 & $2 \times 1.5$\\
 & & & & $2 \times 3.5$\\
Landolt 98 & $06^{h}\,51^{m}\,50^{s}$ & $-00^{\circ}\,21'\,17''$ & Dec 21-24 & $2 \times 1.5$\\
 & & & & $2 \times 3.5$\\
\end{tabular}
\caption{GMOS observing log. The Landolt stars are photometric standards discussed in Landolt (1992).}
\end{table}
\end{center}

The observations are summarised in Table 1.  The field locations
are illustrated in Fig.~\ref{DSS}. The inner field (Field 1) was
placed at the optical edge identified in the Digitized Sky Survey
(DSS) to provide continuity with earlier photographic work.  While
the DSS limit is about 25.5\magarc\ in $B_J$ ($3\sigma$; Corwin
1980), the digitized survey can be displayed at high contrast to
reveal a more extended disk.  The disk edge is broadly consistent
with the photographic limit from photographic amplification (DF 
Malin, personal communication).  The outer field (Field 2) was made
to overlap slightly with Field 1 in order to check the relative
photometric calibration.

The GMOS detector comprises three $2048 \times 4608$ EEV CCDs
arranged in a row where the long dimensions are edge butted
together (Hook\etal\ 2004).
After 2$\times$2 on-chip binning, the pixels are 0.146\arcsec\ in
size.  The square field of view in a single GMOS image is 5.5\arcmin\
on a side.  The on-instrument wavefront sensor and small CCD pixels
provided a corrected psf of 0.60\arcsec\ FWHM for Field 1, and
0.63\arcsec\ FWHM for Field 2. The psf was found to be mildly
elliptic at about the 10$-$15\% level.  A small psf is essential
for removing contaminating sources in the field.  The background
levels between exposures varied by no more than 5\% generally
and the photometric conditions were excellent.

Each of the fields was observed in 9$\times$900 sec dithered
exposures.  This allowed us to remove cosmic ray events and edge
effects arising from the three edge-butted CCDs. Initially, we used
the standard IRAF/Gemini reduction package.  The processing pipeline
includes bias subtraction and flat fielding ({\sl  gireduce}),
mosaicking of individual GMOS CCDs into a single reference frame
({\sl  gmosaic}), followed by combining the 9 dithered exposures
({\sl  imcoadd}).  The images were then trimmed to the area of
common coverage. The final summed images are shown as insets in
Fig.~\ref{DSS}.

The FITS data files arrived without field position information in
the headers. Our next step was to astrometrically calibrate the
summed field images. This was made possible by using the GAIA package
developed by Peter Draper at the University of Durham. We identified
20 stars in each field with known astrometric positions in the
United States Naval Observatory (USNO) photographic catalogue (Monet\etal\
2003).

Once the fields were correctly aligned with respect to each other,
we formed a mask for each field in order to reject unwanted pixels.
Both masks are shown in Fig.~\ref{MASK}. We remove bright foreground stars
($r^{'} < 21$), 
bright galaxies, and pixels which fall in the shadow of the on-instrument
wavefront sensor. We also trim the outer perimeter of both fields where
the SNR falls off due to the dithering process. The proportion of pixels
lost as a function of galactocentric radius is shown in Fig.~\ref{COUNTS}$b$,
discussed below. The star counts are normalized to the dotted curves
in this figure.

\section{Stellar photometry}

Fields 1 and 2 were analyzed using the IRAF {\sl daophot} package
(Stetson 1987; Davis 1994).  For each field, catalogues of $3
\sigma$, $4 \sigma$ and $5 \sigma$ source detections were produced
with the {\sl daofind} routine. For each catalogue,  preliminary
aperture photometry was obtained using the {\sl  phot} task.  The routine
{\sl  psfselect} was used to select approximately the 100 brightest
stars which were then inspected visually. Isolated stars (i.e.
without bright neighbours) which were not saturated were chosen for
psf stars.  There were about 15 such stars in each field and they
were used as input for the task {\sl  psf} which iteratively computes
the point-spread function model for both fields.  Finally, the
task {\sl  allstar} was run to simultaneously fit this model to
stars from the {\sl daofind} catalogues and determine their photometry.
The task {\sl daophot} identified and returned fitted parameters for
15671 and 4468 sources in Fields 1 and 2 respectively; these numbers
dropped to 15390 and 4291 sources after trimming the field edges.

We examined the residual maps produced by {\sl allstar} to evaluate
psf subtraction with different thresholds.  In general, the residuals
were slightly cleaner for Field 1 when compared with Field 2 but
for both fields the residuals were less than 1\% of the total source
flux.  In Fig.~\ref{PSF}, we show a small region from Field 1 before
subtraction, after subtracting sources above $5 \sigma$ and after
subtracting sources above $3 \sigma$.  The excellent photometry and
source subtraction is evident.

The photometric calibrations were established with two Landolt
fields (see Table 1) for which there were half a dozen useful stars
in each field (Landolt 1992). The equivalent magnitude for a single
photoelectron was determined to be $m_{r^{'}} = 28.20$, \ie\ 0.1 mag
fainter than what is quoted at the Gemini GMOS web site. The
photometric standards showed that the photometry was better than
0.04 mag from night to night, and better than 0.02 mag in
consecutive exposures.

In Fig.~\ref{MAGHIST}, we show the magnitude distribution of detected
sources found with {\sl  daophot}. Our catalogue is complete to $r'
= 27.0$ ($3\sigma$), $r' = 26.7$ ($4\sigma$), $r' = 26.3$ ($5\sigma$).
For comparison, the expected GMOS performance can be assessed with
the web-based calculator\footnote{See
www.gemini.edu/sciops/instruments/instrumentITCIndex.html}, with
fair agreement.  Our calculation assumes a G5III spectrum.  For 50
percentile observing conditions, an $R_c=26$ point source in the
$r^\prime$ filter requires 8100 sec (9$\times$15 min dithered
exposures) for a $5\sigma$ detection. The conversion from $R_c$ to
$r'$ is given at the end of this section.

Fig.~\ref{MAGHIST} therefore establishes that the observations were
undertaken in better than 50 percentile conditions.  As a further
check on the integrity of our data, the star counts in the overlap
region between Fields 1 and 2 were compared. We find 10\% more stars
in Field 1 compared to Field 2, which is broadly consistent with
the higher average background and slightly degraded seeing in the
outer field.  As we stressed earlier, the psf subtraction was cleaner
for Field 1 compared to Field 2, and this may contribute to the
difference. The magnitude offset between Field 1 and Field 2
was less than 0.04 mag and consistent with measurements from the
Landolt standards. The uniformity between fields is evident from
Fig.~\ref{MAGHIST}.

It is convenient to establish a correction from $r'$ to $R_c$ for
two reasons. Most surface brightness profiles published to date are
measured in traditional Johnson-Cousins bands (\eg\ Kregel\etal\
2002). Secondly, in a later section, we correct the star counts for
the background contribution due to distant galaxies. Again, galaxy
count surveys to date are quoted in traditional bands (\eg\
Metcalfe\etal\ 2001).  Girardi\etal\ (2002; 2004) derive the requisite
transformation from stellar atmospheric models. In the temperature
range 4000-10,000~K, $r'-R_c$ changes from about 0.3 to 0.15, with
a stronger dependence for cool stars. This compares favourably with
Fukugita\etal\ (1996) who find $r'-R_c = 0.16(V_c-R_c) + 0.13$ such
that for stars with $V_c-R_c$ colours of 0.2 to 0.7, $r'-R_c$ lies
in the range 0.16 to 0.24. Thus, to a good approximation, we can
convert our magnitude system to $R_c$ using $r'-R_c = 0.2$.

\bigskip
\begin{figure}
\caption{
$r'$ magnitude distribution for all sources detected
by {\sl  daophot}. The three curves are from the (top) 3$\sigma$ catalogue,
(middle) 4$\sigma$ catalogue, (bottom) 5$\sigma$ catalogue.
}
\label{MAGHIST}
\end{figure}
\bigskip
\bigskip
\begin{figure*}
\caption{
The derived $r'$ magnitude in {\sl daophot} for all fitted sources 
as a function of major axis radius. The decline in source counts as a 
function of radius is evident. Note the incompleteness at 12\arcmin\ 
due to crowding in the bright inner disk.  Note also the excellent
photometric consistency of the data between Fields 1 and 2.
}
\label{MAGDIST}
\end{figure*}
\bigskip

\bigskip
\begin{figure}
\caption{(a) Source counts per arcmin$^{2}$ as a function
of radius in the outer GMOS fields for (top) $3\sigma$, (middle)
$4\sigma$ and (bottom) $5\sigma$ detections.  The radial binning
is performed in projected elliptic annuli with 100 pix width
(14.6\arcsec). The background galaxy counts have not been removed.
The error bars are given by $1/\sqrt{N}$ where $N$ is the number 
of stars measured within the annular bin.  (b) The upper solid 
histograms show the total number of
CCD pixels per radial bin in Fields 1 and 2. The lower solid curves
show the number of pixels rejected by the mask in Fig.~\ref{MASK}. The dotted
curves show the impact of the rejected pixels on two upper curves.
}
\label{COUNTS}
\end{figure}
\bigskip

\section{Results}

Previous studies have shown that the slope of the radial
light distribution of some disks becomes abruptly steeper at an
outer break radius (\eg\ Pohlen\etal\ 2002; Kregel\etal\ 2003); the
reason for this break is not known.  We now seek to establish whether
the stellar disk truncates rapidly beyond the break or extends
faintly into the HI disk.  Both outcomes have implications for how
most of the baryons must have settled during dissipation.

We derive the radial profile of NGC 300 from the catalogue of fitted
stars. The centre of the galaxy was taken from the NASA/IPAC
Extragalactic Database. Since the disk is inclined, we need to
define elliptic annuli; each annulus has a radial (annular) thickness
of 100 pixels. The form of the annuli is illustrated in Fig.~\ref{MASK}
for which we adopt a PA $=$ 110\deg\ and inclination $i=42$\deg\
(Kim\etal\ 2004).  The angle subtended by our fields on the sky is
33\deg\ with respect to the nucleus; this translates to 43\deg\
projected onto the plane of the disk, or about an eighth of the
total azimuthal extent.  Note that the disk major axis passes through
the middle of Field 1.

In Fig.~\ref{COUNTS}$a$, we show the radial profile in NGC~300
derived from our source counts {\it prior to removal of the background
galaxy counts.} The form of the profile is essentially identical
for the $3\sigma$, $4\sigma$ and $5\sigma$ catalogues. 
The radial profile is normalized by keeping track of the number of
pixels in each annular bin. In Fig.~\ref{COUNTS}$b$, we see that a
typical bin has almost 200,000 pixels. The top hat distributions
are similar in form, and overlap at 18.5\arcmin.  The combined
distribution gives uniform coverage from about 12.5\arcmin\ out to
24.5\arcmin.  The dotted lines indicate the small corrections
required for pixels which are masked out due to bright foreground
stars, bright galaxies and so on (see Fig.~\ref{MASK} and \S 4).
The star counts are normalized to the dotted curves in this figure.

In Fig.~\ref{COUNTS}$a$, there is a gentle transition at 19\arcmin\
(11.4 kpc) where the profile appears to flatten off.  The transition
is somewhat emphasized by the nature of the log-linear axes traditional
for this field.  This happens to fall in the overlap region between
Fields 1 and 2, as we show in Fig.~\ref{COUNTS}$b$.  However, the
change in slope is not an artefact of joining two data sets, since
the form of the transition is identical in both fields. The transition
is an artefact of a faint background galaxy population observed
through the outer disk of NGC~300.

In order to derive the intrinsic surface brightness profile for
NGC~300, we need to remove the background galaxy counts.  In
Fig.~\ref{SUBCOUNTS}, we show three possible levels for background
subtraction.  In Fig.~\ref{SUBCOUNTS}$a$, we take the average galaxy
counts from the Hubble Deep Fields as derived by Metcalfe\etal\
(2001), where we have taken $r'-R_c = 0.2$ (see \S 5).
Fig~\ref{SUBCOUNTS}$b$ takes the outer third of Field 2 to define
the background counts.  This is 10\% higher than the averaged Hubble
Deep Field counts. Given that the expected variance in a 5.5\arcmin\
field down to $r'=27$ can be as high as 50\% from field to field,
this would not be unexpected.  The difference in counts between the
Hubble Deep Fields when integrated down to $r'=27$ is about 15\%
(N. Metcalfe, personal communication).

For Figs.~\ref{SUBCOUNTS}$a$ and $b$, the key implication is that
the disk extends to at least 24\arcmin\ (2.2$R_{25}$), close to 10
optical scale lengths for this low luminosity galaxy.  It is possible
to remove the transition at 18.5\arcmin\ completely with a background
level which is 100\% higher than the Hubble Deep Field counts.  This
outcome is presented in Fig.~\ref{SUBCOUNTS}$c$. In contrast to
Figs.~\ref{SUBCOUNTS}$a$ and $b$, the interpretation here would be
very different.  The disk now appears to truncate very sharply at
18\arcmin, or about 50\% further out in radius than the Holmberg
radius. Such a cut-off corresponds to about 6 optical scale lengths,
although note that the disk is still detected to 1.9$R_{25}$. While
we consider this high background level to be unlikely, it cannot
be ruled out until more of the outer perimeter of NGC~300 has been
mapped under similar observing conditions.

We are not able to separate faint stars from faint background
galaxies using the GMOS images. We attempted to separate these with
{\sl Sextractor} but most of the sources identified as galaxies
could not be distinguished from stars convolved with the mildly elliptic psf.
We therefore make the basic assumption that stars and galaxies have
the same magnitude distribution in the range $r'=23-27$. We can
then directly convert the magnitudes and number counts in
Fig.~\ref{SUBCOUNTS}$b$ to a mean stellar surface brightness as a
function of radius.

In Fig.~\ref{SURFBRI}, photometric data for NGC~300 are superimposed
from three sources: (i) Carignan's photographic measurements in
$B_J$, (ii) $I$ band measurements from Kim\etal\ (2004), and (iii)
our new GMOS $r'$ measurements.  The $r'$ stellar surface brightness profile
is derived from the $3\sigma$ catalogue presented in Fig~\ref{SUBCOUNTS}$b$
after background subtraction.  Note that this profile reaches
exceedingly faint levels of $\mu_{r^{'}} = 30.5$\magarc\ (at $2\sigma$
when averaged over the last two bins in Fig~\ref{SUBCOUNTS}$b$).

We have offset Kim's measurements downwards by 0.5 mag ($r'-I=0.5$),
and Carignan's measurements upwards by 1.1 mag ($B_J-r'=1.1$). Our
population synthesis model in Fig.~\ref{CMD} indicates that an
appropriate correction is $B_J-r'=1.4-1.5$. After applying our
correction for stellar incompleteness below, we cannot account for
the 0.3 mag offset with respect to Carignan's data; this discrepancy
was also noted by Kim\etal\ when they compared their data with
Carignan (1985).

In Fig.~\ref{LOGSURFBRI}, the logarithmic axes emphasize the different
fitted components for the stellar core, the bulge and the disk. The
fitting procedure is described by Kim\etal\ (2004). Roughly speaking,
the data conform to an exponential disk out to 24\arcmin. The
substructure along the curve may be indicative of our restricted
azimuthal coverage.  The bulge cannot be strongly constrained in
Kim's fitting.  However, our figure does make the point that there
is {\it no} compelling evidence for a transition from disk to bulge
or halo stars at the outermost observed extent.

The fraction of missing light was estimated from a theoretical
colour-magnitude diagram derived from the StarFISH code (Harris \&
Zaritsky 2001).  In our models, we explore two possibilities: (i)
an intermediate aged 8 Gyr stellar disk with [Fe/H]=-0.87, consistent
with the metallicity derived from the red giant population
(Tikhonov\etal\ 2005); (ii) an old 12 Gyr disk with metallicity
[Fe/H]=-0.7 comparable to the thick disk of the Galaxy.  We specify
a Salpeter initial mass function and generate 10$^5$ stars for each
model. The mass function uses a (minimum, maximum) stellar mass of
(0.1,100) M$_\odot$, and a binary fraction of 25\%.  The adopted
isochrones are taken from http://www.te.astro.it/BASTI (Pietrinferni\etal\
2004).

The resulting colour-magnitude diagrams are shown in Fig.~\ref{CMD}.
Note that the models predict that the GMOS data recover (i) 54.6\%
(ii) 55.8\% of the total light at our 3$\sigma$ threshold. Therefore,
we estimate our completeness at 55\% in the $r'$ images. This would
indicate that our surface brightness profile in Figs.~\ref{SURFBRI}
and ~\ref{LOGSURFBRI} should be shifted upwards by -0.65 mag in
order to correct for the missing light.  (We do not attempt to
correct for the internal dust extinction.) However, we note that
the galaxy inclination ($i=42$\deg) renders the surface brightness
profile brighter by about -0.32 mag. Therefore, the true upward
correction is expected to be -0.33 mag.

With our derived conversion from $r'$ to $R_c$ (\S 6), this indicates
that the GMOS data reach down to an effective surface brightness
of $\mu_{R_c} = 30.3$\magarc.  The CMD simulations in Fig.~\ref{CMD}
predict an outer disk colour $B-I \approx 1.7$. The equivalent
surface brightness in $B$ reached by the GMOS data is at the level
of $\mu_B = 32$\magarc.  These are levels which cannot be reached
reliably in diffuse light imaging as we discuss in the Appendix.

\bigskip
\begin{figure}
\caption{Star counts as a function of major axis radius for the $3
\sigma$ detections. The counts in the overlap regions have been
averaged. We show the counts after subtracting 3 different background
galaxy count levels indicated by the horizontal line: (a) predicted
counts from averaged Hubble Deep Fields; (b) average source counts
in the outer 8 radial bins (annular width of 800 pixels) assuming
that all sources are galaxies;  (c) an artificially high background
from assuming that all detections in the range 20\arcmin-24\arcmin\
are background galaxies.  In each case, the background level is
shown by a solid line. The error bars are discussed in the caption
for Fig.~6.  
} \label{SUBCOUNTS} \end{figure} \bigskip

\bigskip 
\begin{figure}
\caption{ The observed $r'$ surface brightness profile of NGC 300
derived {\it directly} from the star counts.  The inner crosses are
taken from Kim\etal\ (2004): these are $I$ band measurements shifted
downwards by 0.5 mag.  The open circles are taken from the $B_J$
photograpic measurements of Carignan (1985) and are shifted upwards
by 1.1 mag.  The GMOS data points are derived from the $3 \sigma$
source catalogue presented in Fig.~\ref{SUBCOUNTS}$b$.  After
correcting for incompleteness ($-0.65$ mag) and inclination ($+0.33$
mag), we obtain the intrinsic surface brightness profile (see \S 6);
this amounts to shifting all points upwards by $0.32$ mag.
} 
\label{SURFBRI} \end{figure} \bigskip
\bigskip \begin{figure} 
\caption{ The $r'$ surface brightness profile of NGC 300 presented
in log radius to emphasize the contribution of the different fitted
components.  The data points are described in Fig.~\ref{SURFBRI}.
The continuous curves are the different fitted components, for which
the functional forms are given in Kim\etal\ (2004). The model for the
core is not shown.
} \label{LOGSURFBRI}
\end{figure} \bigskip
\bigskip \begin{figure} 
\caption{The colour-magnitude diagram for a stellar population with 
(left) mean age 8 Gyr and [Fe/H]=-0.9, (right) mean age 12
Gyr and [Fe/H]=-0.7. The intrinsic stellar brightness reached by
our data is indicated by the horizontal dashed lines for 1$\sigma$,
3$\sigma$ and 5$\sigma$ detections. The right hand figure of each
panel shows the
fraction of received light from the stellar population above our
threshold level.  The greyscale indicates the number of stars
used in the simulation. } \label{CMD} \end{figure}

\section{Discussion}


Our data clearly reveal the presence of a faint stellar disk out
to at least a radius of 24\arcmin\ (14.4 kpc) or about 10 disk
scale lengths.  Our data extend far beyond the surface photometry
of Kim\etal\ (2004) and show that the disk remains exponential to
the radial limits of our counts.  There is no evidence for a break
in our data, which cover an angular wedge of about 45\deg\ within
the deprojected  disk -- the stellar surface brightness is well
described by an exponential over almost 10 scale lengths. There is
no evidence of a bulge or a halo component.

Indeed, in three wide-field photometric studies of NGC~300, there
is no compelling evidence for a `break radius' in the galaxy
luminosity profile (de Vaucouleurs \& Page 1962; Carignan 1985;
Kim\etal\ 2004).  Carignan (1985; Fig.~5) traces the light in a
deep UK Schmidt plate out to $\mu_B = 27.5$\magarc\ at $r=12.5$\arcmin.
While there may be a hint at a break in Carignan's Fig.~14 at about
10\arcmin, his earlier figures demonstrate quite different structure
when the galaxy is split into two halves.


Since our stellar counts extend to the outer HI disk, it is important
to consider the possible effects of a warped stellar disk. Puche\etal\
(1990) find evidence for a warp in the HI disk beyond 10\arcmin\
but this does not necessarily indicate that the stellar disk is
warped in the same way, if at all (e.g. Garcia-Ruiz, Sancisi \&
Kuijken 2002). Dramatic stellar warps, which are rarely observed,
are most often associated with strong galaxy interactions.  There
is no obvious interacting companion observed in the vicinity of NGC
300. A weak stellar warp may exist and its effect would be to slow
the radial decline in the stellar luminosity profile.  However,
there is no such effect evident in either Fig.~\ref{SUBCOUNTS} or
Fig.~\ref{LOGSURFBRI}.


The surface brightness of the outer disk of NGC 300 extends smoothly
down to at least 30 $r^{\prime}$\magarc\ which corresponds to a
stellar surface density of order $0.01$ M$_\odot$ pc$^{-2}$. To the
best of our knowledge, such an extended and very diffuse population
of stars has not been seen before.  What is its origin?

A key issue is just how far the HI disk extends.  Puche\etal\ (1990)
have mapped the outer disk in NGC 300 and detect HI out to 32\arcmin,
beyond the extent of the newly established optical disk.  The
HI surface density profile is roughly constant at $0.5-1$ M$_\odot$
pc$^{-2}$ in the outer disk. The surface density profile of the
optical disk falls with radius, and is roughly comparable to the
HI at the Holmberg radius, assuming $M/L_R = 1$. Beyond here, the
optical surface density falls by a factor of 100 at a radius of
20\arcmin\ or more.

The existence of stars at such low surface densities is difficult to
understand in the conventional picture. How does a spiral density wave
initiate and propagate -- if this is indeed what triggered star
formation in the outer disk -- in such a rarefied medium?  

The $Q$ criterion is often used to establish whether a disk is
unstable to axisymmetric modes, viz.
\begin{equation} Q = {{\sigma
\kappa}\over{3.36 G \Sigma}} 
\end{equation}
where $\sigma$ is the
internal dispersion of the HI gas which stabilizes the disk on small
scales (up to the radius of the Jean's mass). On larger scales, the
disk is stabilized by differential rotation which is embodied by
$\kappa$, the epicyclic frequency of the disk. In contrast, the
disk is destabilized by its own surface density, $\Sigma$. The disk
is locally stable against axisymmetric modes if $Q$ is substantially
above unity.

We derive $\kappa(r)$ and $\Sigma(r)$ from the data of Puche\etal\
(1990) and assume $\sigma = 5$\kms\ in the outer disk. We thus find
that $Q$ is everywhere greater than 5$\pm$2 beyond 10\arcmin\ which
would appear to argue against star formation and the existence of
spiral density waves here.

However, the observations of Puche\etal\ (1990) not only show that
the HI extends well beyond the extent of our stellar disk but also
reveal well-defined, tightly wound spiral structure in the HI which
is particularly prominent in the SE at 20\arcmin.  While this spiral
structure is difficult to explain, it is interesting that HI spiral
arms are increasingly observed in the outer parts of disk galaxies
(TA Oosterloo, personal communication; Bureau\etal\ 1999; Quillen
\& Pickering 1997). Bureau \etal\ found a similar problem for the
HI disk of NGC 2915, which also shows marked spiral structure at
very large radius; they invoked the effect of a tumbling triaxial
halo potential to drive the spiral structure in this apparently
stable outer disk.


It is known from photometric imaging studies (e.g. de Jong 1996)
that most Hubble types have colour gradients from the centre outwards
which arise from (i) the decreasing dust content; (ii) declining
stellar abundances; and (iii) changing stellar populations due to
radial variations in star formation history. The general trend is
that disks become progressively bluer at larger radii and this has
been ascribed to the stellar population becoming younger on average
and progressively more metal poor (de Jong 1996).

However, the inference that the stellar populations become progressively
younger with radius contrasts with what we know from recent stellar
photometry of the outermost reaches of nearby galaxies.  These
studies incorporate high quality stellar photometry and show, by
comparison with well calibrated globular cluster data, that the
outer giant populations are metal poor and old.  Recent examples
include the outer disks of M33 (Galleti, Bellazzini \& Ferraro 2004)
and M31 (Ferguson \& Johnson 2001).  

The data for NGC 300 are limited at present but Tikhonov\etal\
(2005) find that metal poor ([Fe/H]=-0.87) giants dominate the light
at large radius and these are most likely to be old. We note that
Tikhonov\etal\ attribute various regions of NGC 300 to the thick
disk or to the halo which is difficult to do in a relatively face-on
galaxy. One can really only rely on indirect arguments based on our
Galaxy, but these are very unreliable. For example, their outermost
field (S3) corresponds to a deprojected radius of about 11 kpc or
roughly 6 scale lengths.  In our Galaxy, this corresponds to about
20 kpc where we know nothing about the stellar disk at the present
time.


So is the outer disk in NGC 300 young or old? When one considers
the effects of changing extinction and metallicity, a moderately
blue outer disk does not necessarily indicate a younger stellar
population.  (Galleti\etal\ (2004) discuss how age and metallicity
effects can be disentangled in outer disk/halo observations.) Within
the context of cold dark matter models, the answer to this question
has key implications for the formation scenario for the outer disk
(e.g.  Ferguson \& Johnson 2001).

If follow up measurements reveal that the outer disk is moderately
young, we would expect to find evidence of spiral density waves in
the stellar population even at the implied large radii ($>$12 kpc),
although these spiral arms may be moderately diffuse and hard to
recognize.  As mentioned above, there is clear evidence of spirality
in the outer HI disk.  If stars are forming today, it may be that
the outer HI disk largely comprises cold, compact clouds with
internal dispersions lower ($\lta 1$\kms) than one normally associates
with beam-averaged observations of outer disks ($\sim 5$\kms).  If
the HI gas was to cool to 100 K, and nonthermal (e.g.  turbulent)
velocities were $\lta 1$\kms, the $Q$ criterion could be as low as
unity.  Such clouds would be gravitationally unstable today and the
outer disk would have a moderately young stellar component.  But
we stress that, to our knowledge, the HI velocity dispersion has
not been measured in distinct clouds in the outer disk of NGC 300.

If new measurements reveal that the outer disk is old, then the
current high value of $Q$ here may simply reflect that the HI has
been largely used up today, but that there was a time $\sim$10 Gyr
ago $(z\sim 1)$ when there was far more gas than we observe today.
If our assumption of the cloud dispersion velocity still holds,
this would have the effect of raising the surface density of the
disk, and lowering the $Q$ parameter to the point where local
instabilities induce star formation.


Hydrogen in the outermost reaches of galaxies may be evidence for
gas accretion, either from gas settling onto the dark halo (cold
accretion), or from a galaxy merger event (hot accretion).  If the
material in the slow warp in NGC 300 is the result of gas accretion
to the outer disk, it is plausible that the settling process produced
low levels of star formation. An example of such a process is
dramatically illustrated by the gas disk settling within the giant
elliptical NGC~5128 where a young blue population is clearly evident
(Graham 1979; Bland, Taylor \& Atherton 1987).

But how does an accretion process succeed in maintaining the
continuity of the exponential disk? This is particularly puzzling
when the surface brightness distribution is contrasted with that
of most other spirals, which show a clear break in their light
distribution at 3 to 5 radial scale lengths. Could the Lin-Pringle
viscous evolution of a star-forming disk explain the persistence
of the exponential disk of NGC 300 to ten or more scale lengths?

Another possibility is that the stars were scattered from the inner
disk. N-body simulations of disk evolution indicate that radial
mixing is strong (Sellwood 2002). A single spiral wave near co-rotation
can perturb the angular momentum of a star by $\sim$20\% moving it
either inwards or outwards by several kiloparsecs (Sellwood \&
Kosowsky 2002). This process could help to wash out any stellar
disk edge but it is difficult to see how this explains the very
extensive stellar disk, with a simple exponential density profile,
observed here.

\acknowledgments
JBH would like to dedicate this work to his brother Simon who died
in December 2004.  We thank the staff of the Gemini Observatory for
their tireless efforts and their pursuit of excellence. MV was
supported by an AAO Student Fellowship in the austral winter of 2004.
BTD was supported in part by NSF grant AST-9988126.  We are
grateful to Laura Dunn for her help with running the StarFISH code,
and to the referee for a careful examination of this paper.

\newpage
\references
Barton IJ \& Thompson LA 1997, AJ 114, 655 \\
Battaner et al 2002, A\&A 388 213 \\
Bender R 1990, A\&A 229 441 \\
Bernstein R\etal\ 2002, ApJ 571 56 \\
Blair WP \& Long KS 1997, ApJS 108 261 \\
Bland J, Taylor K \& Atherton PD 1997, MNRAS 228, 595 \\
Brown TM\etal\ 2003, ApJ 592 L17 \\
Bureau M, Freeman KC, Pfizner DW \& Meurer GR 1999, AJ 118 2158 \\
Butler DJ, Martinez-Delgado D \& Brandner W 2004, AJ 127 1472 \\
Carignan C 1985, ApJS 58 107 \\
Corwin HG 1980, MNRAS, 191, 1 \\
Davis LE 1994, A Reference Guide to the IRAF/DAOPHOT Package \\
de Grijs R\etal\ 2001, MNRAS 324 1074 \\
de Jong RS 1996, A\&A, 313, 377 \\
de Vaucouleurs G \& Capaccioli M. 1979, ApJS 40 699 \\
de Vaucouleurs G \& Page J, 1962, ApJ 136, 107 \\
Draine BT 2003, ApJ 598 1017 \\
Durrell PR\etal\ 2001, AJ 121 2557 \\
Ferguson AMN \& Clarke CJ 2001, MNRAS 325, 781 \\
Ferguson AMN \& Johnson RA 2001, ApJ 559, L13 \\
Freedman W \etal \ 2001, ApJ 553 47 \\
Freeman KC 1970, ApJ 160 811 \\
Freeman KC \& Bland-Hawthorn J 2002, ARAA 40 489 \\
Fukugita \etal \ 1996, AJ 111 1748 \\
Galleti S, Bellazzini M \& Ferraro FR 2004, A\&A 423, 925 \\
Garcia-Ruiz I, Sancisi R \& Kuijken K 2002, A\&A 394, 769 \\
Gieren W \etal \ 2004, AJ 128 1167 \\
Girardi L 2002, A\&A 391 195 \\
Girardi L 2004, A\&A 422 205 \\
Graham JA 1979, ApJ 232 60 \\
Haikala LK \& Mattila K 1995, ApJ 443 L33 \\
Harris J \& Zaritsky D 2001, ApJS 136 25\\
Hook IM\etal\ 2004, PASP 116 425 \\
Jerjen H, Freeman KC \& Binggeli B 1998, AJ 116, 2873 \\
Kennicutt R 1989, ApJ 344 685 \\
Kim SC \etal \ 2004, ChJAA 4 299 \\
Kregel M, van der Kruit PC \& de Grijs R 2002, MNRAS 334, 646 \\
Labb\'{e}, I.\etal\ 2003, ApJ 591, L95 \\
Landolt A 1992, AJ 104 340 \\
Li A \& Draine BT 2001, ApJ 554, 778 \\
Li A \& Draine BT 2002, ApJ 576 762 \\
Lin DNC \& Pringle JE 1987, ApJL 320, L87 \\
Maloney PR \& Bland-Hawthorn J 2001, ApJ 553 L129 \\
Metcalfe \etal \ 2001, MNRAS 323 795 \\
Monet D\etal \ 2003, AJ 125 984 \\
Pietrinferni A\etal\ 2004, ApJ, 612, 168\\
Pohlen M\etal\ 2002, A\&A 392 807 \\
Pohlen M\etal\ 2004, astro-ph/0405541 \\
Pritchet CJ \& van den Bergh S. 1994, AJ 107 1730 \\
Puche D, Carignan C, \& Bosma A. 1990, AJ 100, 1468 \\
Quillen AC \& Pickering TE 1997, AJ, 113, 2075\\
Reynolds RJ 1992, ApJ 392 L35 \\
Roberts WW 1969, 158, 123 \\
Sakai S\etal\ 1997, ApJ 478 49 \\
Seiden P\etal\ 1984, ApJ 282 95 \\
Sellwood JA 2002, In Disks of Galaxies: Kinematics, Dynamics \& Perturbations, ed E Athanassoula, A Bosma, R Mujica, 275, 281 (San Francisco: PASP) \\
Sellwood JA \& Kosowsky A 2002, In The Dynamics, Structure \& History of Galaxies, ed G Da Costa, H Jerjen, 273, 243 (San Francisco: PASP) \\
Schlegel DJ, Finkbeiner DP \& Davis M 1997, AAS 191 8704 \\
Shu F, Milione V \& Roberts WW 1973, ApJ 183 819 \\
Stetson PB 1987, PASP 99 191 \\
Szomoru A \& Guhathakurta P 1998, ApJ 494 L93 \\
Tikhonov NA, Galazutdinova OA \& Drozdovsky IO 2005, A\&A 431, 127 \\ 
van der Kruit P 1979, A\&AS 38 15 \\
van der Kruit P 1987, A\&A 173 59 \\
Weiner BJ\etal\  2001, ApJ, 546, 916 \\
Weingartner JC \& Draine BT 2001, ApJ 548 296 \\
Witt, AN, \& Vijh, UP 2004, in Astrophysics of Dust, ed. AN Witt, G.C. Clayton, \& BT Draine 
   (San Francisco: Astr. Soc. Pacific), p. 115.\\

\newpage
\appendix
\section{Starlight scattered by dust in the outer disk}

In our study, we reach an effective surface brightness of $\mu_{r^{'}}
= 30.5$\magarc\ ($2\sigma$).  This level is exceedingly faint in
the sense that the equivalent surface brightness in $B$ or $V$ is
fainter than 32\magarc\ for an old stellar population.  This raises
an important question: Is there a practical limit to galaxy photometry?
Here, we demonstrate that in principle the metal-poor outer HI
envelope has a sufficient column of dust to scatter detectable
levels of starlight from the optical disk.  Thus, we propose that
only star counts can reliably trace luminosity profiles in galaxies
at this faint level or even fainter.

Let us consider a disk with inclination angle $\incl$.  Starlight
from the inner regions of the galaxy will illuminate the dust grains,
resulting in scattering.  Consider a region in the galaxy where the
H nucleon column density normal to the disk is $N_{\rm H}$.  If we
approximate the illuminating starlight as due to a point source at
a distance $D$ with luminosity $L_\nu$ located at the galactic
center, then the dust grains at a radial distance $r$ from the
galactic center will result in scattered intensity $I_\nu$, where
\beq \label{eq:I_nu} I_\nu = \frac{L_\nu}{4\pi r^2} \frac{N_{\rm
H}}{\cos\incl} .  F_X (\lambda,\Theta) \eeq The scattering phase
function can be written \beq F_X (\lambda,\Theta) \equiv \sum_j
\frac{n_j}{n_{\rm H}} \left(\frac{d\sigma}{d\Omega}\right)_{j,\lambda}
\eeq for which $(d\sigma/d\Omega)_j$ is the differential scattering
cross section for grain type $j$, at wavelength $\lambda$ and
scattering angle $\Theta_s$, and $n_j/n_{\rm H}$ is the number of
grains of type $j$ per H nucleon.  Eq. (\ref{eq:I_nu}) can be
rewritten
\beq
\label{eq:Imagarcsec2}
\frac{I_{\nu}}{29\ {\rm mag}~ {\rm arcsec}^{-2}}
=
\frac{1}{\cos\incl}
\left( \frac{r/D} {\rm arcsec} \right)^{-2} 
\left( \frac{N_{\rm H}} {10^{18}\cm^{-2}} \right)
\left( \frac{F_X(\lambda,\Theta_s)} {10^{-24}\cm^2\sr^{-1}} \right)
10^{0.4(14-m_{\rm gal})}
\eeq
where $m_{\rm gal}$ is the apparent magnitude of the galaxy. For
NGC~300, the NED data base gives $m_{\rm gal} = 8.95$ in the $B$
band rising to $m_{\rm gal} = 7.5$ in the $R$ band.

\begin{figure*}[h]
\caption{$F(\lambda,\Theta)$ or Milky Way dust with $R_V=3.1$ 
        (lower panel) and
	SMC bar dust with $R_V=2.9$ (lower panel) at
	selected wavelengths $\lambda$ (SDSS z', i', r', g', and u',
	and Cousins I$_c$, R$_c$, and V$_c$), as a function of the
	scattering angle $\Theta_s$.
	}
	\label{PHASEFUNC}
\end{figure*}
\begin{figure*}[h]
\caption{Degree of polarization for scattering by Milky Way dust with
        $R_V=3.1$, and
	SMC bar dust with $R_V=2.9$ at
	selected wavelengths $\lambda$, as a function of the
	scattering angle $\Theta_s$.
	}
	\label{POL}
\end{figure*}

If the location in the galaxy is specified by the observed position
angle $\psi$ (measured from the minor axis, with $\psi=0$ corresponding
to the side of the galaxy nearer to us) and displacement $\Theta$
from the center, then

\beq
\frac{I_{\nu}}{29\ {\rm mag}~ {\rm arcsec}^{-2}}
=
\frac{\cos\incl(1+\tan^2\psi)}{1+\cos^2\incl\tan^2\psi}
\left( \frac{\Theta} {\rm arcsec} \right)^{-2} 
\left( \frac{N_{\rm H}} {10^{18}\cm^{-2}} \right)
\left( \frac{F_X(\lambda,\Theta_s)} {10^{-24}\cm^2\sr^{-1}} \right)
10^{0.4(14-m_{\rm gal})}
\eeq
The scattering angle $\Theta_s$ is related to inclination $\incl$ and
position angle $\psi$ by
\beq
\Theta_s = \arccos 
	\left[ 
		\frac{\sin\incl}{\sqrt{1+\cos^2\incl\tan^2\psi}}
	\right]
\eeq
leading to $\Theta_s=90^\circ$ for a face-on galaxy ($i=0$). 

Weingartner \& Draine (2001) have developed dust models that reproduce
the observed wavelength-dependent extinction in the Milky Way, LMC,
and SMC.  These models are also consistent with the observed infrared
and far-infrared emission (Li \& Draine 2001, 2002).  The scattering
properties of these dust models have been calculated by Draine
(2003).  Fig. 11 shows $F(\Theta_s)$ from Draine (2003) at selected
wavelengths for MW dust and SMC dust.  Note that $F(\lambda,\Theta_s)$
for the SMC is lower than for the Milky Way by approximately a
factor 10, which is approximately the metallicity of the SMC relative
to solar.  Thus, it seems reasonable to estimate
$F_X(\lambda,\Theta_s)\approx (Z/Z_{\sun})F_{\rm MW}(\lambda,\Theta_s)$
as an estimate for $F_X$ in another galaxy.

Puche et al.\ (1990) measured $N({\rm H\,I})\approx 9\times10^{19}\
{\rm cm}^{-2}$ in NGC 300 at 1000\arcsec.  For the outer disk of
NGC 300, the red giant colour distribution indicates [Fe/H] $=$
-0.87 (Tikhonov, Galazutdinova \& Drozdovsky 2005); we take
$Z/Z_{\sun}\approx 0.2$ for the gas.  Evaluating eq. (\ref{eq:Imagarcsec2})
for a point on the major axis 1000\arcsec\ from the center, we find
$I \approx 31.5\ {\rm mag}\ {\rm arcsec}^{-2}$, or about 4\% of the
stellar surface brightness at this point.  Scattered starlight makes
a small contribution to the total surface brightness for NGC 300
at $r=1000$\arcsec, but for other cases (e.g., galaxies where the
stellar surface density declines more rapidly) scattered starlight
could contribute a larger fraction of the total surface brightness.

Scattered starlight, if present, will be substantially polarized.
In Fig.~\ref{POL}, we show the degree of linear polarization expected
for the scattered light as a function of scattering angle, where
we assume that the galactic starlight incident on the dust grains
is unpolarized.  For a $90^\circ$ scattering angle (e.g. for a
face-on galaxy) the polarization is expected to be $\sim45$\% in
red light, with smaller values at shorter wavelengths.

In addition to scattering light, dust grains appear to luminesce
in the red when illuminated by ultraviolet light (see Witt \& Vijh
2004, and references therein).  This so-called ``Extended Red
Emission'' can be comparable in intensity to the scattered starlight
in the red.

\end{document}